\documentclass[pra,superscriptaddress,showpacs,amsmath,amssymb,twocolumn]{revtex4}
\usepackage{bm}
\usepackage{graphicx}

\begin{document}

\newcommand{\tr}{\mathop{\mathrm{tr}}\nolimits}
\newcommand{\adj}{\mathop{\mathrm{adj}}}
\newcommand{\diag}{\mathop{\mathrm{diag}}}
\renewcommand{\Re}{\mathop{\mathrm{Re}}}
\renewcommand{\Im}{\mathop{\mathrm{Im}}}

\newcommand{\R}{\mathop{\mathbb{R}}\nolimits}

\newcommand{\SA}{\mathop{\cal{S}}\nolimits}
\newcommand{\E}{\mathop{\cal{E}}\nolimits}
\newcommand{\M}{\mathop{\cal{M}}\nolimits}

\newtheorem{Thm}{Theorem} \newtheorem{Prop}{Proposition} \newtheorem{Lem}{Lemma} \newtheorem{Cor}{Corollary} 
\newtheorem{Def}{Definition} 

\newcommand{\bracket}[2]{\langle #1 | #2 \rangle}
\newcommand{\ketbra}[2]{| #1 \rangle \langle #2 |}
\newcommand{\bra}[1]{\langle #1 |}
\newcommand{\ket}[1]{| #1 \rangle}

\title{Entropies in General Probabilistic Theories and its Application to Holevo Bound}
\author{Gen Kimura}
\email{gen@shibaura-it.ac.jp}
\author{Junji Ishiguro}
\affiliation{Shibaura Institute of Technology, Saitama, 337-8570, Japan}
\author{Makoto Fukui}
\affiliation{Keio University, Tokyo, 108-8345 Japan}
\date[]{Feb. 24, 2016}
\begin{abstract}
General probabilistic theories are designed to provide operationally the most general probabilistic models including both classical and quantum theories. 
In this letter, we introduce a systematic method to construct a series of entropies, all of which generalize Shannon entropy in classical system and von Neumann entropy in quantum system. 
Using these entropies, the Holevo bound, an upper bound of the accessible information from a quantum system, is generalized to hold in any general probabilistic theory. 
\end{abstract}


\maketitle

\section{Introduction}

One of the key concept to construct the modern theory of information is a probability. 
Shannon's information theory is based on the classical probability theory \cite{ref:Shannon}, while the quantum information theory is based on the quantum probability theory \cite{ref:NC}. 
However, classical and quantum probability theories are not only theories that provide operationally valid probabilistic models, but there are infinitely many such other probabilistic models. 
General probabilistic theories (hereafter GPTs) are designed to provide all of the operationally valid probabilistic models (See for instance \cite{ref:Mackey,ref:Araki,ref:Gudder1,ref:Ludwig}). 
In recent years, many researchers are trying to construct the general information theories based on GPTs \cite{ref:Barrett1,ref:Barrett,ref:BHK,ref:CDP,ref:Hein,ref:SW,ref:BE,ref:KNI10,ref:BBKM15}. 
Through this line of research, it is mainly expected to understand the nature of information processings without resort to not only classical concepts (i.e., local realism) but also particular rules of quantum theory. 
For instance, it is possible to construct a safe key-distribution protocol which is based on experimentally testable physical principles, such as the no-signaling condition and the existence of an entanglement \cite{ref:BHK}. 

To quantify information, the concept of entropy plays an important role.  
The optimal compression ratio is given by Shannon entropy in classical system and by von Neumann entropy (hereafter vN entropy) in quantum system, respectively. 
Entropy can be also used to characterize an upper bound of the accessible information \cite{ref:Holevo}: From a quantum system, the accessible information $I(\{p_x,\rho_x\})$ of encoded state $\rho_x$ each prepared with a probability $p_x$ is bounded by  
\begin{equation}\label{eq:HolevoChi}
I(\{p_x,\rho_x\}) \le S_q(\rho) - \sum_x p_x S_q(\rho_x), 
\end{equation}
where $\rho = \sum_x p_x \rho_x$ and $S_q(\rho):= -\tr \rho \log \rho$ is the vN entropy. (In this letter, we use the binary logarithm).  
In constructing the general information theory, therefore, it would be useful to have the concept of entropy in each GPT.  
So far, at least three entropies $S_1,S_2,S_3$ (See Eq.~\eqref{eq:S1S2S3} below) are known \cite{ref:Hein,ref:SW,ref:BE,ref:KNI10}, all of which are well-defined irrespective of the underlining probabilistic models (See also \cite{ref:BBKM15}). 
Those quantities are defined as suitable positive-semidefinite functionals on a state space of GPT, such that they coincide with Shannon entropy and vN entropy if the model is classical and quantum, respectively. 
Interestingly, they differ from each other in general GPT, only degenerating when a model is classical or quantum. 
However, as for the application of information gain, it is known that none of them in general provide an upper bound of the accessible information in the form of Eq.~\eqref{eq:HolevoChi} \footnote{Since $S_2$ and $S_3$ are not concave in general, the right hand side of \eqref{eq:HolevoChi} can be negative; $S_1$ is concave but a counter example of \eqref{eq:HolevoChi} has been given in \cite{ref:BE}.}.  
In this letter, we introduce the novel method to construct new entropies which are {\it induced} from a known entropy in a way that they still generalize both Shannon and vN entropies. 
We show that the combination of the original entropy and the induced entropy gives an upper bound of the accessible information in any GPT. 
Since the von Numenann entropy is {\it invariant} under the induction, this result generalizes the Holevo theorem to hold in any GPT. 
We exemplify these results in the squared model, which is the simplest GPT neither classical nor quantum. 
In this model, we show that $S_1,S_2,S_3$ are interrelated through the induction: The induced entropy from $S_1$ coincides with $S_2$ and the induced entropies from $S_2$ and $S_3$ coincide and are invariant under the induction.  

This letter is organized as follows. 
In Sec.~\ref{sec:Entropies}, we introduce the induced and invariant entropies investigating their general properties.  
In Sec.~\ref{sec:HolevoBdd}, we generalize the Holevo theorem to hold in any GPT.  
These results are exemplified in the squared model in Sec.~\ref{sec:Squared}.  
Finally, in Sec.~\ref{sec:Conc}, we conclude this letter with several future works.

\section{Entropies in GPTs}\label{sec:Entropies}

In this letter, we assume the reader's familiarity with GPTs but give the short review mainly for the consensus of notations. 
(For the details of GPTs, we refer for instance \cite{ref:Barrett,ref:KNI10} and references therein). 
In any GPT, a physical state can be represented by a vector such that a probabilistic mixture of states corresponds to a convex combination of the represented vectors. 
Hence, the set of all states $\SA$ is a convex set in the underling vector space. 
A state is called pure if it cannot be prepared by a non-trivial probabilistic mixture; otherwise called mixed. 
Geometrically, a state is pure iff it is an extreme point of $\SA$. 
With a given state $s \in \SA$, ${\cal D}(s)$ (resp. ${\cal P}(s)$) is the set of probabilistic mixture $\{p_x,s_x\}$ such that $s = \sum_x p_x s_x$ (resp. with $s_x$ begin pure).

An affine functional $e: \SA \to [0,1]$ is called an {\it effect}, and any measurement (with a finite outcome) can be represented by a tuple of effects $M = (m_y)_y$ such that 
$m_y(s)$ gives the probability to get $y$th outcome when the measurement is performed under a state $s$. 
We denote by $\E, \M$ and $\M_{\mathrm{fg}}$ the sets of all effects, measurements and fine-grained measurements \footnote{
A measurement $M = (m_j)_j$ is called fine-grained \cite{ref:SW} if any $m_j$ are indecomposable. An non-zero effect $e$ is called indecomposable \cite{ref:KNI10} if $e = e_1 + e_2 \ (e_1,e_2 \in \E)$ then there exists $c_1,c_2 \ge 0$ such that $e_i = c_i e \ (i=1,2)$.}, respectively.  

In any GPT, three entropies are defined through classical-information quantities as follows \cite{ref:Hein,ref:SW,ref:BE,ref:KNI10}: 
\begin{subequations}\label{eq:S1S2S3}
\begin{align}
S_1(s) &:= \inf_{M = (m_y)_y \in \M_{\mathrm{fg}}} H(m_y(s)), \label{eq:S1}\\
S_2(s) &:= \sup_{\{p_x,s_x\} \in {\cal P}(s) }\sup_{M = (m_y)_y \in \M_{\mathrm{fg}}} I(X:Y), \label{eq:S2} \\
S_3(s) &:= \inf_{\{p_x,s_x\} \in {\cal P}(s)} H(p_x). \label{eq:S3} 
\end{align}
\end{subequations}
Here, $H(p_i):= -\sum_i p_i \log p_i$ is the Shannon entropy and $I(X:Y): = H(X)+H(Y)-H(X,Y)$ is the mutual information between random variable $X$ and $Y$. 
As is mentioned above, all $S_1,S_2,S_3$ coincide with Shannon entropy or vN entropy when the model is classical or quantum, respectively, but they differ in general. 

Here, we introduce a method to construct a new entropy from these known entropies: 
\begin{Def}\label{def:IndS}
With $S$ being any entropy of GPT, we define an {\it induced entropy} $S'$ by  
\begin{equation}\label{eq:ind}
S'(s):=\sup_{\bigl\{p_x,s_x \bigr\} \in {\cal D}(s)} \bigl\{\sup_{(m_y)\in\ \M_{\mathrm{fg}}} I(X:Y) +\sum_x p_x S(s_x)\ \bigr\}
\end{equation}
\end{Def}
Note that the range of measurement $\M_{\rm fg}$ can be enlarged to the whole measurement set $\M$ since any measurement can be decomposed into a fine-grained measurement \cite{ref:KNI10} while a mutual information does not decrease through the decomposition by the data process inequality \cite{ref:CT}. 
Note however that the range of decomposition ${\cal D}(s)$ cannot be restricted to ${\cal P}(s)$ (See footnote [20]). 

Interestingly, both Shannon and vN entropies are invariant under induction \eqref{eq:ind};  
In other words, we have:
\begin{Thm}\label{thm:GenvNEnt}
Let $S_q$ and $H$ be vN and Shannon entropies, then $S'_q = S_q$ and $H'= H$.  
\end{Thm}
{\bf Proof.} 
By definition, $S_q'$ reads 
$$
S_q'(\rho):=\sup_{\bigl\{p_x,\rho_x \bigr\} \in {\cal D}(\rho)} \bigl\{\sup_{(m_y)\in\ \M} I(X:Y) +\sum_x p_x S_q(\rho_x)\ \bigr\}
$$
where $\rho$ is a density operator. 
By Holevo theorem \cite{ref:Holevo}, we have $I(X:Y) + \sum_x p_x S_q(\rho_x) \le S_q(\rho) $ for any preparation $\{p_x,\rho_x\} \in D(\rho)$ and for any POVM measurement $M$.
Thus we have $S'_q(\rho) \le S_q(\rho)$. 
Next, with a preparation $\bigl\{p_x,\rho_x \bigr\} = \bigl\{p_x,\ketbra{\phi_x}{\phi_x}\bigr\} $ and a POVM measurement $(m_y) = (\ketbra{\phi_y}{\phi_y})$ where 
$\rho = \sum_x p_x \ketbra{\phi_x}{\phi_x}$ is an eigenvalue decomposition of $\rho$, we have $I(X:Y) = H(Y) - \sum_x p_x H(Y|X=x) = H(\bracket{\phi_y}{\rho \phi_y}) - \sum_x p_x H(\bra{\phi_y}\ketbra{\phi_x}{\phi_x} \ket{\phi_y}) = H((p_y)) = S_q(\rho)$, 
and $\sum_x p_x S_q(\rho_x) = 0$ since $S_q$ vanishes on pure states.  
Thus, we obtain $S_q (\rho) = I(X:Y) + \sum_x p_x S_q(\rho_x) \le S_q'(\rho)$. 

The invariance of $H$ can also be shown using the fact that a classical model is embedded into a quantum model (using only diagonal elements).
\hfill $\blacksquare$ 


Notice that Theorem \ref{thm:GenvNEnt} implies that if $S$ is a generalization of vN (resp. Shannon) entropy in quantum (resp. classical) system, 
then, so is the induced entropy $S'$. 
Therefore, starting from such entropy, e.g., $S_1,S_2,$ or $S_3$, the induction \eqref{eq:ind} provides a systematic method to construct a series of entropies in any GPT which generalize both Shannon and vN entropies. 

We shall call an entropy $S$ in a GPT an {\it invariant entropy} if it is invariant under induction \eqref{eq:ind}: $S'(s) = S(s) \ \forall s \in \SA$. 
Both Shannon and vN entropy are thus examples of an invariant entropy. 
If $S$ is an invariant entropy, then for any state $s \in \SA$ and for any decomposition $\{p_x,s_x\} \in {\cal D}(s)$, it follows that $S(s) = S'(s) \ge \sup_{(m_y)\in\ \M} I(X:Y) +\sum_x p_x S_q(\rho_x) \ge \sum_x p_x S_q(\rho_x)$. 
Namely, we have proved:
\begin{Prop}
If $S$ is invariant, then is concave. 
\end{Prop}

Notice that it is highly non-trivial whether there exists an invariant entropy in arbitrary GPT. 
However, the following argument suggests the affirmative answer about the existence: 
Firstly, note that the induction is generally non-decreasing since a preparation of $s$ with probability $1$ is one of a preparation of $s$ and the fact $I(X:J) = 0$ for such deterministic preparation:
\begin{Prop}\label{prop:NDofInd}
$S'(s) \ge S(s) \ (\forall s \in \SA)$. 
\end{Prop}
Therefore, we have infinitely many series of non-decreasing entropies $S \to S' \to S'' \to \cdots $ through induction \eqref{eq:ind}.
Secondly, if GPT is finite, i.e., if the dimension of an underlying vector space is finite, then it is not difficult to show that these sequence are bounded by above which is independent of a state, 
hence there exists the limit for each state. 
We shall call this an {\it infinity entropy}, and it is a natural conjecture that an infinity entropy is invariant. 
We later come back to this problem in Sec.~\ref{sec:Squared} where the existence is shown in a squared model. 

Before going to the application of the induced entropy, let us investigate the role of entropies as a measure of mixedness:  

\begin{Prop}\label{prop:MP} (i) If $S'(s) = 0$, then $s$ is a pure state. (ii) Conversely, if $S$ is an entropy such that $S(s) = 0$ for any pure state $s$, then
 the induced entropy $S'$ also satisfies the property.  
\end{Prop}

{\bf Proof.} 
(i) This can be proven similarly as Proposition 23 in \cite{ref:KNI10}. 
(ii) Let $s$ be a pure state. Then, there exists the unique preparation $\{1,s\}$ and we have $
S'(s) = \sup_{M \in \M} I(X:Y) + S(s) = 0$ since $S(s) = 0$ and $I(X:Y) = 0$ for a deterministic preparation of $X$. \hfill $\blacksquare$ 

We say that an entropy $S$ measures a mixedness (of a state) if $S(s) = 0 \Leftrightarrow s$ is pure. 
Proposition \ref{prop:MP} implies the following: 
\begin{Cor}
If $S$ measures a mixedness, so does $S'$. 
\end{Cor}
In \cite{ref:KNI10}, we have shown that both $S_2$ and $S_3$ measure a mixedness (but not $S_1$ in general), hence any induced entropies from $S_2$ and $S_3$ measure a mixedness of a state.  

\section{Bound on Accessible Information}\label{sec:HolevoBdd}

In this section, we provide an application of the induced and invariant entropy in GPTs. 
Let us start from the general setting of an information gain in a GPT:
Assume that Alice has an information resource $\{p_x,x\}$ preparing a message $x$ with a probability $p_x$, and decodes it to a state $s_x$ in a GPT. 
After sending the state (through a noiseless channel) to Bob, he try to encode the message $x$ by performing a suitable measurement $M = (m_y)_y$. 
The accessible information is defined by the maximum of a mutual information $I(X:Y)$ between a message $x$ and a measurement outcome $y$: 
\begin{equation}
I(\{p_x,x\}) := \sup_{M \in {\cal M}} I(X:Y). 
\end{equation}
Now, we show that the combination of an original entropy $S$ and its induction $S'$ provide an upper bound of an accessible information. 
\begin{Thm}\label{thm:main}
For any encoding $\{p_x, s_x\}$ in arbitrary GPT, the accessible information is bounded by
\begin{equation}\label{eq:HolevoGddGPT}
I(\{p_x,s_x\}) \le S'(s) - \sum_x p_x S(s_x),
\end{equation}
where $s = \sum_x p_x s_x$.
\end{Thm}
{\bf Proof.} 
The proof is almost straightforward by reminding that $\M_{\rm fg}$ can be enlarged to $\M$ in \eqref{eq:ind}: 
By the definition of an induced entropy, we have $I(X:Y) + \sum_x p_x S(s_x) \le S'(s)$ for any encoding $\{p_x,s_x\}$ and measurement $M \in {\cal M}$, 
which completes the proof. 
\hfill $\blacksquare$

Notice that this result (combined with Theorem \ref{thm:GenvNEnt}) generalizes the famous Holevo theorem \eqref{eq:HolevoChi} in quantum system to hold in arbitrary GPT. 
It is worth mentioning that the difficulty to show Holevo theorem lies not in obtaining the upper bound \eqref{eq:HolevoGddGPT} 
but rather in showing the invariance of vN entropy (Note that we have used Holevo's result to show Theorem \ref{thm:GenvNEnt}). 
In particular, if we use an invariant entropy $S$, then the upper bound takes the same form as in \eqref{eq:HolevoChi}.

\section{Squared Model}\label{sec:Squared}
In this section, we illustrate our results in the squared model:  
The state space of the squared model can be represented by $\SA = \{ (c_1,c_2) \in \R^2 \ | \ 0 \le c_i \le 1 \ (i=1,2)\}$. 
There are four pure states $s^{(00)} = (0,0),s^{(01)} = (0,1),s^{(10)} = (1,0),s^{(11)} = (1,1)$. 
Using Table 1 in \cite{ref:KNI10}, without loss of generality, any fine-grained measurement can be parametrized by one parameter $\alpha \in [0,1]$ such that $(m_j(s))_{j=1}^4 = (\alpha c_1, \alpha \bar{c_1}, \bar{\alpha}c_2,\bar{\alpha}\bar{c_2})$ for $s = (c_1,c_2) \in \SA$. Here we use the notation $\bar{a}:= 1- a$ for $a \in [0,1]$. 
Then, for a fine-grained measurement, Shannon entropy reads $H((m_j(s))) = h(\alpha) + \alpha h(c_1) + \bar{\alpha} h(c_2)$ where $h(x)  = - x \log x - \bar{x} \log \bar{x}$ is the binary entropy. 
Moreover, with a fixed decomposition $\{p_x,s_x = (c_{1x},c_{2x})\} \in {\cal D}(s)$, we have $
I(X:J) = \alpha h(c_1) + \bar{\alpha} h(c_2) - \sum_x p_x(\alpha h(c_{1x}) + \bar{\alpha} h(c_{2x}))$. 
Noting the affinity of $\alpha$, we have $S_1(s) = \inf_{M \in \M_{\rm fg}} H((m_j(s))) = \min [h(c_1),h(c_2)]$ \cite{ref:KNI10}, and  
\begin{equation}\label{eq:optI}
\sup_{M \in \M_{\rm fg}} I(X:J) = \max_{i=1,2} [h(c_i) - \sum_x p_x h(c_{ix})].  
\end{equation}
Moreover, since $c_{ix} = 0$ or $1$ for any pure state $s_x$, we have $S_2(s) = \sup_{M \in \M_{\rm fg}} I(X:J) = \max [h(c_1),h(c_2)]$ \cite{ref:KNI10}. 

\begin{Prop}\label{Prop:SqRel}
In the squared model, we have 
$$
S_1(s) \le S'_1(s) = S_2(s) \le S_3(s) \le S'_2(s) = S'_3(s) = S''_2(s),  
$$
implying that $S'_2$ and $S'_3$ are invariant in the squared model. 
For $s = (c_1,c_2)$, their analytic form is given by 
\begin{equation}\label{eq:AFS2ind}
S'_2(s) = S'_3(s) = h(c_1) + h(c_2). 
\end{equation}
\end{Prop}
Interestingly, generally non-relating entropies $S_1,S_2$ and $S_3$ are all connected through our induction in the squared model. 
Since $S_2'$ and $S_3'$ are invariant, this coincides with the infinity entropies and thus the conjecture on the existence of an invariant entropy in Sec.~\ref{sec:Entropies} is satisfied. 

\begin{Cor}
An induced entropy is not necessary concave even from concave entropy. Conversely, induced entropy can be concave from a non-concave entropy. 
\end{Cor}
Indeed, in the squared model, $S_1$ is concave but $S_2(s) = S_1'(s) $ is not concave (See \cite{ref:KNI10}). 
On the other hand, $S_2'(s) = h(c_1) + h(c_2)$ is concave since it is an invariant entropy (alternatively, since $h$ is concave). 

\bigskip

\noindent {\bf Proof of Prop.~\ref{Prop:SqRel}.} 
The inequalities $S_1(s) \le S_2(s) \le S_3(s)$ have been shown in \cite{ref:KNI10}.  
In the following, we denote $s=(c_1,c_2)$ and $\{p_x,s_x = (c_{1x},c_{2x})\} \in {\cal D}(s)$, 
and use Eq.~\eqref{eq:optI} for the estimation of induction \eqref{eq:ind}.   

Noting that $S_1(s) = \min [h(c_1),h(c_2)]$, $S_2(s) = \max [h(c_1),h(c_2)]$, we have $S'_1(s) = \max_{i=1,2} [h(c_i) + \sum_x p_x( \min [h(c_{1x}),h(c_{2x})] - h(c_{ix}))] \le \max_{i=1,2} [h(c_i)] = S_2(s)$ since $\min[a,b] -a \le 0$. 
Using the decomposition $s = p_1 s_1 + p_2 s_2 $ where $s_1 = (1,c_2)$ and $s_2 = (0,c_2)$ with $p_1 = c_1, p_2 = \bar{c_1}$, we have 
$\max_{i=1,2}[h(c_i) + \sum_x p_x(\min [h(c_{1x}),h(c_{2x})] - h(c_{ix}))] = h(c_1)$. 
Using another decomposition $s = p'_1 s'_1 + p'_2 s'_2$ where $s'_1 = (c_1,1)$ and $s'_2=(c_1,0)$ with $p'_1 = c_2, p'_2 = \bar{c_2}$, we have $\sup_{M \in \M_{\rm fg}} I(X:J) - \sum_x p'_x S_1(s) = h(c_2)$. 
Therefore, the above inequality is attainable, and thus $S_1'(s) = S_2(s) = \max [h(c_1),h(c_2)]$. 

Similarly, $S'_2(s) = \max_{i=1,2} [h(c_i) + \sum_x p_x(\max [h(c_{1x}),h(c_{2x})] - h(c_{ix}))]$. 
If $i=1,2$ attains the maximum, this is less than or equal to $h(c_{i}) + \sum_x p_x h(c_{_{j}x}) \le h(c_1) + h(c_2)$ where $j \neq i$ and we have used $\max[a,b] -a \le b$ and the concavity of $h$. 
Moreover, it is easy to see that the inequality is attainable by the above decomposition $s = p_1 s_1 + p_2 s_2$ \footnote{\label{ft:test}
Note that if we restrict the decomposition to pure states for the induced entropy from $S_2$, we have $\sup_{M \in \M_{\rm fg}} I(X:J) = \max_{i=1,2} [h(c_i)] = S_2(s) \neq S'_2(s)$. 
This shows that, different from the case for $S_2$, the optimization over state decomposition in \eqref{eq:ind} cannot be restricted to pure state decomposition. } and thus $S'_2(s) = h(c_1) + h(c_2)$. 

Consider now the pure state decomposition $s =  \sum_x p_x s_x = c_1 c_2 s^{(10)} + c_1 \bar{c_2} s^{(11)} + \bar{c_1} c_2 s^{(01)} + \bar{c_1} \bar{c_2} s^{(00)}$. 
Since $(p_x)$ is a product of $(c_1,\bar{c_1})$ and $(c_2,\bar{c_2})$, we have $H((p_x)) = h(c_1) + h(c_2)$ using the additivity of Shannon entropy. 
By definition \eqref{eq:S3}, we have $S_3(s) \le S'_2(s)$. 
It also follows that \footnote{
For $i=1$, we have $h(c_1) - \sum_x p_x h(c_{1x}) + \sum_x p_x S_3(s_x) \le h(c_1) + h(c_2) \Leftrightarrow \sum_x p_x S_3(s_x) \le h(c_2) + \sum_x p_x h(c_{1x})$, 
but this follows from $S_3 \le S'_2$ and the concavity of $h$. Case $i=2$ follows in the parallel manner.} 
$S'_3(s) = \max_{i=1,2}[h(c_i) - \sum_x p_x h(c_{ix})] + \sum_x p_x S_3(s_x) \le h(c_1) + h(c_2) = S'_2(s)$. 
However, since $S_2(s) \le S_3(s)$, we have $S_3'(s) = S'_2(s)$.  

Finally, we show the invariance of $S'_2$ (and also $S'_3$). 
By Proposition \ref{prop:NDofInd}, we have $S'_2(s) \le S''_2(s)$. 
However, using $S'_2(s) = h(c_1) + h(c_2)$, we have $S''_2(s) = \max_{i=1,2} [h(c_i) + \sum_x p_x ( h(c_{1x}) + h(c_{2x}) - h(c_{ix}))]$;  
if $i = 1,2$ attains the maximum, this equals $h(c_i) + \sum_x p_x h(c_{jx})$ where $j \neq i$, which is less than or equal to $h(c_1) + h(c_2) = S'_2(s)$ by the concavity of $h$. 
This completes the proof. 
\hfill $\blacksquare$

\vspace{-5mm}

\section{Conclusions and Discussion}\label{sec:Conc}

In this paper, we have proposed a systematic method to induce infinitely many entropies in any GPT starting from a well defined entropy such as $S_1,S_2$ and $S_3$. 
In particular, the induction keeps the generalization of Shannon and vN entropy in classical and quantum system and also the property as the measure of mixedness. 
Using the combination of the induced entropy and the original entropy, we have generalized the Holevo theorem to hold in any GPT. 
In a fixed GPT, an invariant entropy seems to play an important role; 
it is always concave and gives an upper bound of the accessible information exactly in the same form as \eqref{eq:HolevoChi}. 
Moreover, Shannon entropy and vN entropy are both invariant in a classical and quantum system, respectively. 
The existence of an invariant entropy in a general model is strongly suggested by the existence of an infinity entropy in any (finite) GPT and is exemplified in the squared model by $S_2'$. 

It would be further interesting to investigate an operational meaning of each entropies (such as the optimal compression ratio), the tightness of the bound, and the relation with the thermodynamical entropy, etc.  


\bigskip 

{\bf Acknowledgement} 
We would like to thank Prof. K. Matsumoto, Dr. K. Matsuura, Dr. K. Nuida for fruitful advices and comments. 


\end{document}